\documentstyle[aps,epsfig,twocolumn]{revtex}
\begin{document}
\draft
\flushbottom
\twocolumn[
\hsize\textwidth\columnwidth\hsize\csname @twocolumnfalse\endcsname


\title{Quantum Spectra and Wave Functions in Terms of Periodic Orbits for Weakly
Chaotic Systems}
\author{R. E. Prange$^1$, R. Narevich$^2$, and Oleg Zaitsev$^1$
\\
$^1$Department of Physics, University of Maryland, \\
College Park, Maryland 20742\\
$^2$Department of Physics and Astronomy, University of Kentucky, \\
Lexington, Kentucky 40506}
\maketitle

\tightenlines
\widetext
\advance\leftskip by 57pt
\advance\rightskip by 57pt

\begin{abstract}
Special quantum states exist which are quasiclassical quantizations of
regions of phase space that are weakly chaotic. In a weakly chaotic region,
the orbits are quite regular and remain in the region for some time before
escaping and manifesting possible chaotic behavior. Such phase space regions
are characterized as being close to periodic orbits of an integrable
reference system. The states are often rather striking, and can be
concentrated in spatial regions. This leads to possible phenomena. We review
some methods we have introduced to characterize such regions and find
analytic formulas for the special states and their energies.

\smallskip\ 

PACS number(s): 05.45.-a, 03.65.Sq, 72.15.Rn

\end{abstract}

\vspace{7mm}
]
\narrowtext
\tightenlines

\section{Special quantum states}

There are many {\em special quantum eigenstates}, [more precisely {\em %
classes} of eigenstates] which often exist in the ``quantum chaos'' problems
that have been considered. These states are associated with special regions
of phase space which are `weakly chaotic'. By this we mean that classical
orbits are regular and remain in the region for a sufficiently long time. It
may be that almost all the orbits are chaotic, but that is a long time
property of little consequence to the formation of quantum states. The
regions can be considered as being close to a set of nonisolated periodic
orbits. These periodic orbits are not necessarily orbits of the system under
study. They could rather be orbits of a `nearby' integrable reference system.

This paper reviews some recent work by the authors on this topic which is
partially published\cite{PNZ},\cite{NPZmag}, and partially submitted for
publication\cite{NPZPhysaE},\cite{NPZFP}. Some further examples and details
will appear in the thesis of Oleg Zaitsev\cite{ZThesis}. We have introduced
a new technique which is in some ways more general than those previously
used. We have also shown how to generalize an older method, the
Born-Oppenheimer approximation, to new situations.

Remarkably, only two such kinds of states have been clearly recognized in
the literature, but there are numerous others which have been overlooked.
This is in spite of the fact that the quantum problems giving rise to these
states have been extensively studied numerically. Even in cases where the
states dominate the effect under consideration they have not been mentioned.
The reason for this oversight is probably as follows: a) the special states
are high quantum number states and they are rare in the sense that they are
a small fraction of the ordinary states existing in the same energy range,
and b) most work has concentrated on energy levels and energy level
statistics rather than wavefunctions.

The one special state known to most physicists and engineers, even those not
especially concerned with `quantum chaos', is Lord Rayleigh's {\em %
whispering gallery }mode\cite{LR}. Although in most respects this case is
very well understood, we shall mention a new development here, as well as
providing a new [or rather old] easy way to get the main result.

There are many phenomena associated with whispering gallery modes. The
whispering gallery in St. Paul's cathedral in London is perhaps the first
case to be remarked by a scientist. In 1871, G. B. Airy attempted to explain
the phenomenon. Lord Rayleigh, in his {\em Theory of Sound }of 1894
disagreed with Airy and gave, at the level of ray analysis, the currently
accepted explanation. It is amusing that the leading order formula taking
the wave nature of sound into account is expressed in terms of Airy
functions. The most recent mention of these modes that we have found is by
Agam and Altshuler\cite{AA}. They have explained the {\em lack} of
whispering gallery modes in the `Faraday' experiments of Kudrolli, et. al.%
\cite{KAG} which shakes a container of water in the shape of a Bunimovich
billiard, as due to the fact that significant dissipation occurs at the
boundary,{\em \ }precisely where the whispering gallery modes are localized.

The second class of states is familiar to most quantum chaologists if not to
the wider physics community. These are the {\em bouncing ball} modes. There
are some phenomena associated with these modes, as well\cite{HM}.

The most widespread class of `bouncing ball' states occurs in mixed chaotic
systems, that is, systems which have stable periodic orbits as well as
unstable ones. Keller and Rubinow\cite{Keller} considered, in particular, a
smooth convex billiard which must in general have a minimum diameter where
there is an orbit bouncing perpendicularly to the boundary at two opposing
points. Such an orbit is stable and there is a volume of phase space near
the orbit which is invariant under the Hamiltonian flow. It's possible to
make a harmonic expansion and quantize these states. More generally, there
are generalizations of such `bouncing ball' states to the neighborhood of 
{\em any }stable periodic orbit. This requires $\hbar $ to be small enough
that the area of the invariant phase space on a surface of section
perpendicular to the orbit is of order Planck's constant $h.$

Although our technique allows an improvement on the harmonic expansion, we
shall not consider further such states. Since they typically occur in mixed
chaotic systems they have received relatively little attention from the
quantum chaos community until recently.

However, there is a rather spectacular generalization of these bouncing ball
states to some hard chaotic systems not possessing stable orbits. The best
known are the modes of the Bunimovich stadium corresponding to a ball
bouncing perpendicularly to the straight sides of the billiard. These were
discovered by MacDonald and Kaufman\cite{MKauf} in their pioneering
numerical work on that system. Similar modes are known in other billiards
with parallel sides\cite{stadium},\cite{biswas}.

These bouncing ball states have been studied from many points of view,
including periodic orbit theory\cite{tanner},\cite{stadSlant}.
Paradoxically, much work has been aimed at `getting rid of them'. This is
because the classical bouncing ball periodic orbits dominate the `trace
formula' and the corresponding quantum states dominate the oscillatory part
of the density of states, thus obscuring the underlying `quantum chaos' of
the energy level correlations, which was deemed more interesting.

Although these special states are associated with periodic orbits, they are
not to be confused with `scars'\cite{scars}. Scarred states are certainly of
great interest. They are associated with unstable periodic orbits, and are a
relatively weak effect. The special states are associated with a class of
reference periodic orbits and are a large effect. They are strongly
localized in some sense or another, and for this reason have the potential
to lead to distinct phenomena. In fact, sometimes one of a sequence of
special states can be regarded as being scarred.

The classical limit of the systems under consideration may be strongly
chaotic, or mixed chaos, or pseudointegrable, or even, if you wish,
integrable. These are {\em long time }characterizations of the classical
motion, however, and have no direct correlation with quantum states. To
support a special state, the system must be weakly chaotic in the sense that
there is region of phase space where the classical motion is nearly
integrable for a short but long enough time. The size of this region and the
characteristic times are functions of $\hbar .$ There is also a condition on
the way that orbits escape from this region.

\section{Some examples}

Here are some examples of systems with special states which have not been
remarked in the literature. We consider only two dimensional billiards, for
simplicity. We begin with three variations on the Bunimovich stadium, which
has horizontal straight parallel sides of length $2a$ capped by semicircles
of radius $R,$ where $a/R$ is of order unity. We consider the solutions of
the Helmholtz equation $\left( \nabla ^2+k^2\right) \Psi (x,y)=0,$ where $%
\Psi $ vanishes on the boundary for values of $k^2$ on the spectrum. We
consider only cases where $k$ is large, but not too large. That is, we do
not automatically take the limit $k\rightarrow \infty $ since the states of
interest will be of negligible number in that limit. This is equivalent to
considering $\hbar $ to be small.

\subsection{Bunimovich stadium}

First, we review the original Bunimovich stadium which has a set of {\em %
nonisolated }bouncing ball orbits, i.e. those orbits bouncing
perpendicularly between the straight sides. Such a set of orbits gives an
especially large contribution to the Gutzwiller trace formula, as is well
known and we will remind you below. This leads to the result that there is a
very strong {\em correlation} {\em of energy levels. }In fact, the energy
level correlation is dominated by the {\em \ }bouncing ball{\em \ quantum
states}. These are states of the approximate {\em Born-Oppenheimer }form, 
\begin{equation}
\Psi _{nm}(x,y)=\Phi _n(y|x)\psi _m(x),  \label{BunPsi}
\end{equation}
where 
\begin{equation}
\Phi _n(y|x)=\sin n\pi \frac{y+R-\xi _B\left( x\right) }{2\left( R-\xi
_B\left( x\right) \right) }.  \label{Phixy}
\end{equation}
[The Born-Oppenheimer treatment of the Bunimovich billiard was first given
in reference \cite{Taylor} and is discussed further in reference\cite
{stadium}.] Here $\xi _B\left( x\right) =0,$ $\left| x\right| \leq a,$ $\xi
_B\left( x\right) =R-\sqrt{R^2-\left( \left| x\right| -a\right) ^2}\approx
\frac 12\left( \left| x\right| -a\right) ^2/R$ for $\left| x\right| \geq a,$ 
$\left( \left| x\right| -a\right) <<R.$ This makes $\Phi (y|x)$ vanish when $%
y$ is on the billiard boundary. Treating $x$ as a slowly varying parameter
in $\Phi (y|x),$ it is seen that $\psi _m(x)$ satisfies 
\begin{equation}
-\psi _m^{\prime \prime }(x)+V(x)\psi (x)=E_m\psi (x)  \label{Spsi}
\end{equation}
where $V(x)$ is a sort of square well potential, 
\begin{equation}
V(x)\approx n^2\pi ^2\xi _B(x)/2R^3  \label{V(x)}
\end{equation}
and the energy 
\begin{equation}
k^2=k_{n,m}^2\approx \left( n\pi /2R\right) ^2+E_{m.\text{ }}  \label{energy}
\end{equation}
Thus, for large $n,$ $\psi _m(x)$ is something like $\sin (m\pi (x+a)/2a)$,
with $m<<n.$ Therefore $E_m\approx (m\pi /2a)^2.$

The leading term of the energy, $\left( n\pi /2R\right) ^2$, is a
quantization of the bouncing ball orbits. The next term, is a quantization
of the transverse motion, motion in the potential $V(x).$ If this
quantization is carried out semiclassically, it too is expressed in terms of
the periodic orbits of motion in the potential.

These states have wavenumbers $k_{n,m}\approx n\pi /2R+O(1).$ This
corresponds to the leading term in the trace formula, coming from the
nonisolated bouncing ball orbits, which varies as $\exp (4ikR).$ The length $%
s_1=4R$ satisfies, approximately, $\exp (is_1(k_{n+N,m}-k_{n,m}))=1.$ It is
the leading element of the {\em length spectrum. }The next element, $s_2=8R$
also satisfies this condition. In the trace formula it comes from once
repeated bouncing ball orbits. Thus the bouncing ball states account the
most of the correlations in the length spectrum which are multiples of $4R.$

We show in Fig. 1 a density plot of $\Psi _{10,1},$ as given by Eq. (\ref
{BunPsi}). [Actually, in most cases we show the absolute square of the
theoretical wavefunction.] The wavefunctions we show are theoretical. Later
we make some remarks on the accuracy of the theoretical predictions.

\vspace{2mm}
\begin{figure}[tbp]
{\hspace*{0.2cm}\psfig{figure=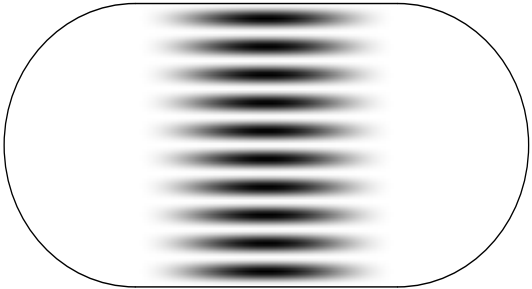,height=4.4cm,width=7.5cm,angle=0}}  
{\vspace*{.13in}}
\caption[ty]{Bouncing ball mode of the standard Bunimovich stadium. Straight side
has length $2a,$ endcap radius is $R=a.$ The quantum numbers are $n=10,$ $%
m=1.$ Figs. 1,2,3,4,6,7 give the stadium shape and a density plot of the
square of the theoretiacl wavefunction.}
\end{figure}   

It is, of course, not true that the bouncing ball orbits by themselves give
the bouncing ball states\cite{tanner}. These orbits occupy a set of measure
zero in phase space. Nevertheless, the most natural way to characterize the
phase space which supports the states is that it is in a certain
neighborhood of the bouncing ball periodic orbits.

\subsection{Slightly sloped stadium}

The first variation is the slightly sloped stadium considered by Primak and
Smilansky\cite{stadSlant}. In this case the end caps have radius $R(1\pm
\epsilon )$ and the sides, instead of being horizontal, have slopes $\pm
\epsilon R/a.$ The parameter $\epsilon $ is small. [To minimize diffraction,
we suppose that the sides merge smoothly to the end caps with no change of
slope. The results are given to leading order in $\epsilon $ only.] In this
case, the special states are {\em strongly modified } even if $\epsilon $ is
quite small.

In fact, the formulas above hold upon replacing $\xi _B\rightarrow \xi _S$
where the major new feature is that, for $\left| x\right| <a,$ 
\begin{equation}
\xi _S\approx -\epsilon xR/a.  \label{xiS}
\end{equation}
This gives for $V(x)$ a steep sided well with a sloping bottom, 
\begin{equation}
V(x)\approx -\epsilon n^2\pi ^2x/2aR^2,  \label{VSlope}
\end{equation}
for $\left| x\right| <a.$ The condition that the slope be large enough to
change substantially the states from the Bunimovich, $\epsilon =0,$ case is $%
\epsilon n^2/R^2\geq 1/a^2.$ If this inequality is strong, the small $m$
states $\psi _m$ are approximately Airy functions concentrated near $x=a.$
In Fig. 2 we show the state $\Psi _{10,2}$ for $\epsilon =0.05,$ $a=R.$

\vspace{2mm}
\begin{figure}[tbp]
{\hspace*{1.2cm}\psfig{figure=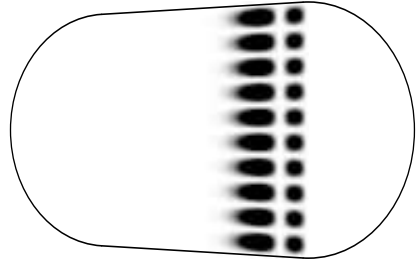,height=4.5cm,width=6.0cm,angle=0}}  
{\vspace*{.13in}}
\caption[ty]{Special mode of a slightly sloped Bunimovich stadium. The slope
parameter is $\epsilon =0.05,$ and $R=a.$ The quantum numbers are $n=10,m=2.$}
\end{figure}   

One motivation for considering this case is that the nonisolated periodic
orbits mathematically disappear for any finite $\epsilon $ no matter how
small. However, if $\epsilon $ is small, there is a region of phase space
which remains close to the bouncing ball orbits of the original stadium.
Thus, there are still special states related to the bouncing ball orbits of
the standard stadium. However, it is complicated and not very enlightening
to describe these states in terms of the periodic orbits of the slightly
sloped stadium itself.

There is a parametrically different dependence on $\epsilon $ of the states
as compared with the energy correlations. We see below that if $kR\epsilon
<<1,$ the leading terms of the trace formula, or in other words, the energy
correlations due to the bouncing ball states, are little modified\cite
{stadSlant}. This is the natural expectation, since the condition is simply
that the wavelength is large compared with the change of radius of the
endcap. On the other hand, we have just found that if $ka\sqrt{\epsilon }%
\geq 1$ the eigenstates and energies are substantially changed. Thus, if $%
\left( ka\right) ^{-2}<<\epsilon <<(kR)^{-1}$, the states are changed but
the correlations are not.

\subsection{Baseball stadium}

As $\epsilon $ in the previous example becomes of order unity, the special
state again changes. Note that for finite $\epsilon ,$ there is a set of
nonisolated periodic orbits through the center of the larger circle, whose
angular range is of order $\epsilon .$ If the endcaps have radii $R_2$, $R_1$
with $0<R_2-R_1<2a,$ the angular range of such periodic orbits is $\pm
\theta _a$ measured away from the vertical, where $\sin \theta _a=\left(
R_2-R_1\right) /2a.$ This region supports special states $\Psi
_{n,m}(r,\theta )$, where $r,$ $\theta $ are polar coordinates for the large
circle.

The Born-Oppenheimer approximation used above relies on having a {\em fast}
variable and a {\em slow} variable. The ansatz, Eq. (\ref{BunPsi}), was made
with $y$ the fast variable and $x$ the slow. Fast and slow mean from the
classical point of view that the motion in the $y$ direction is much faster
than in the $x$ direction. From a wave viewpoint it means that $\left|
\partial \Phi /\partial y\right| >>\left| \partial \Phi /\partial x\right| $
which allows the $x$ dependence of $\Phi $ to be treated parametrically.

The states of this section would seem to have a fast radial motion and a
slow angular motion. However, close to the center of the circle this
separation is not valid. Nevertheless, the Born-Oppenheimer approximate
state gives the result for the outer region, and that suffices.

Thus, the states are approximated as above, replacing the fast coordinate $%
y\rightarrow r$ and the slow coordinate $x\rightarrow \theta .$ Then 
\begin{equation}
\Phi (r|\theta )\approx \frac 1{\sqrt{kr}}\cos \left( kr+\frac{\nu (\theta
)^2-\frac 14}{2kr}+\alpha (\theta )\right)  \label{Phir}
\end{equation}
for $kr$ large compared with $\nu $. This function is an asymptotic Bessel
function of order $\nu $, [a superposition of Bessel and Neumann functions]
and $\psi (\theta )$ satisfies $\psi ^{\prime \prime }(\theta )+\nu (\theta
)^2\psi (\theta )=0.$ The billiard boundary is $r_\partial (\theta )=R_2+\xi
_C(\theta )$ and $\Phi $ must vanish for $r=r_\partial (\theta ).$ Here $\xi
_C=0,$ $-\theta _a<\theta <\pi +\theta _a$ and $\xi _C=R_2(1/\cos (\theta
+\theta _a)-1)\approx \frac 12R_2(\theta +\theta _a)^2$ for $\theta <-\theta
_a$ and $(\theta +\theta _a)^2<<1,$ with a similar formula near $\theta =\pi
+\theta _a.$ It can be shown that $\alpha =\frac 12k(\xi _C(\theta +\pi
)-\xi _C(\theta ))\pm \pi /4$ and $\nu ^2=E_m-V(\theta )$ with $V(\theta
)=k^2R_2(\xi _C(\theta +\pi )+\xi _C(\theta )).$ For large enough $k$ there
is an approximate solution $\psi (\theta )=\sin (m\pi (\theta +\theta
_a)/2\theta _a),$ $\left| \theta \right| <\theta _a,$ $\psi (\theta )$
vanishes outside this region for $\left| \theta \right| <\pi ,$ and $\psi
(\theta )=\pm \psi (\theta +\pi ).$ For this solution $E_m=\left( m\pi
/2\theta _a\right) ^2.$ The energy parameter $k\,$ solves $%
kR_2+E_m/kR_2=(n\mp \frac 14)\pi .$ For large $n$ this has the approximate
solution $k_{n,m}^2=\left( (n\mp \frac 14)\pi /R_2\right) ^2-E_m/R_2^2.\,$
There are $2n-\frac 12\mp \frac 12$ radial nodes.

We show this state for $n=10,$ $m=$ 1, in Fig. 3. Notice some changes of
sign as compared with the bouncing ball between the straight sides. In
particular, a deviation of the billiard sides in the direction of {\em %
narrowing} the channel gives a repulsive effective potential $V$, thus
containing the wave function. For a billiard which is a deviation from a
circle, an outward deviation gives the repulsive effective potential.

\vspace{2mm}
\begin{figure}[tbp]
{\hspace*{0.2cm}\psfig{figure=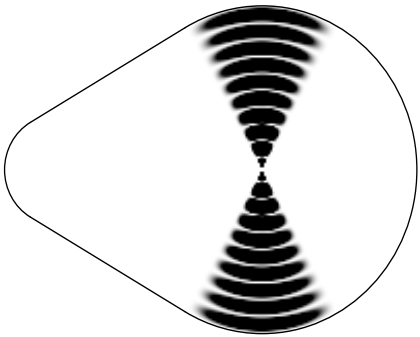,height=4.8cm,width=6.0cm,angle=0}}  
{\vspace*{.13in}}
\caption[ty]{Special mode of a `baseball' stadium. $R_2=1.5a,$ $R_1=\,0.5a.$
Quantum numbers are $n=10,$ $m=1.$ The theoretical wave function in this and
the next figure is inaccurate near the center of the circle.}
\end{figure}   

\subsection{Almost circular stadium}

The above Eq. (\ref{Phir}) can be applied to find the low angular momentum
states of any almost circular billiard. Energy levels and wavefunction
statistics were extensively studied in this system\cite{circle}. Take $%
r_\partial =R+\xi _D(\theta )\,$to describe the billiard, with $\xi
_D(\theta )$ small. For the Bunimovich stadium with side $a<<R$ , $\xi
_D\approx a\left| \cos \theta \right| $. Polar coordinates at the center of
the billiard are used and approximations neglecting $\left( a/R\right) ^2$
have been made.

In this case the potential $V(\theta )=2k^2aR\left| \cos \theta \right| ,$
an attractive triangular well near $\theta =\pm \frac 12\pi .$ It is
periodic with period $\pi .$ For large $k^2aR,$ the lowest eigenstates will
be concentrated near $\theta =\pm \frac 12\pi .$ We show the state $n=10,$ $%
m=2$ in Fig. 4.

\vspace{2mm}
\begin{figure}[tbp]
{\hspace*{1.4cm}\psfig{figure=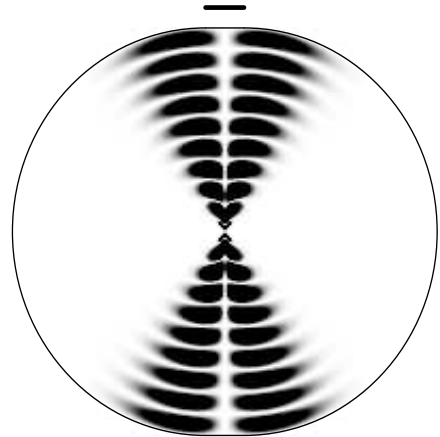,height=6.2cm,width=6.0cm,angle=0}}  
{\vspace*{.13in}}
\caption[ty]{Special low angular momentum mode in a Bunimovich stadium with a
short side, $a=0.1R.$ Quantum numbers are $n=10,$ $m=2.$}
\end{figure}   

All of these variations on the Bunimovich stadium have been considered in
the literature, except for the baseball stadium. The only special states
previously pointed out, however, were for the Bunimovich stadium itself.

\section{Finding and approximating special states}

\subsection{History}

The whispering gallery modes and the bouncing ball modes corresponding to
stable periodic orbits were first quantized by Keller and Rubinow\cite
{Keller} by what is sometimes called the {\em ray method. }This technique
starts with the classical mechanics or rays, and exploits caustics and
adiabatic invariants. However, the assumptions usually made are
unnecessarily strong, and these special states still exist even if the
caustics are only a short time approximation.

Another group of methods starts from the partial differential equation, the
Helmholtz equation in the case of billiards. The {\em parabolic equation
method }invented independently by Leontovich\cite{Leon} and by Fock\cite
{Fock} chooses coordinates astutely, and finds appropriate scale factors,
allowing an approximation to the partial differential equation. It often
relies on the ray method to motivate the manipulation of the PDE. The {\em %
etalon }method of Babich and Buldyrev\cite{BB} is an improvement of this
which involves choosing a characteristic example, or `etalon', which
captures the essential features of the given problem. The {\em extended
Born-Oppenheimer method }[EBO] has been used in this context only for
bouncing ball states between parallel sides of a billiard. We showed above a
couple of generalizations of this technique. Other generalizations are given
elsewhere\cite{NPZFP}.

The last three methods are closely related. They use different language and
motivation and make approximations in a different order, but the main result
is the same. Systematic corrections to the leading result have been
extensively studied for certain examples, and these corrections are
apparently of a different form for the different methods. We prefer the EBO
when it applies since it is simpler and better known than the other
techniques.

\subsection{Bogomolny operator}

We also have introduced a technique based on Bogomolny's {\em surface of
section transfer operator}\cite{bogolss} which in certain ways is more
general than the above methods. It also has the advantage that the transfer
operator $T$ is closely related to the trace formulas and to important
`resummations' of the trace formulas. In addition, it makes somewhat more
precise the notion of a region of phase space which is nearly integrable for
short times.

The operator $T(s,s^{\prime }|E)$ introduced by Bogomolny is a
generalization of the {\em boundary integral method} used to obtain
numerical solutions for billiard problems. The main equation of the boundary
integral method is derived using Green's theorem. It takes the form

\begin{equation}
\mu (s)=\int_\partial ds^{\prime }K(s,s^{\prime }|E)\mu (s^{\prime }),
\label{BIM}
\end{equation}
a Fredholm integral equation. The kernel or operator $K$ depends
parametrically on the energy $E.$ Eq. (\ref{BIM}) has a nontrivial solution
only when $E$ is on the spectrum. For Dirichlet conditions, $\mu
(s)=\partial \Psi ({\bf r)}/\partial n,$ the normal derivative of the
wavefunction evaluated at the position on the boundary labelled by $s$ the
distance along the perimeter. The integral in Eq. (\ref{BIM}) is over this
boundary.

The boundary coordinate $s$ together with its conjugate momentum, the
momentum parallel to the boundary $p_s,$ gives the Birkhoff coordinates for
a surface of section of a billiard. Bogomolny's method essentially
approximates $K(s,s^{\prime }|E)$ by its asymptotic form, which we call $%
T(s,s^{\prime }|E).$ It generalizes to a broad class of surfaces of section,
not restricted to billiards, for which there are often also exact kernels.
For a billiard, in Birkhoff coordinates, 
\begin{equation}
T(s,s^{\prime }|E)=\left( \frac{k\left| \partial ^2L(s,s^{\prime })/\partial
s\partial s^{\prime }\right| }{2\pi i}\right) ^{\frac 12}e^{i\left(
kL(s,s^{\prime })+\mu \right) },  \label{Tdef}
\end{equation}
where $L$ is the distance between boundary points, $k=\sqrt{2mE}/\hbar $ and 
$\mu =\pi $ for Dirichlet conditions. We shall suppress $\mu $ and write the
prefactor as $\left( .\right) $ for simplicity. The essential variation of $%
T $ is given by the exponential dependence on $kL=\hbar kL/\hbar
=S(s,s^{\prime }|E)/\hbar ,$where $S$ is the action of the classical
straight line orbit between boundary points $s,s^{\prime }.$

Because $\hbar $ is supposed to be small compared with typical classical
actions the exponential is rapidly varying, and the method of stationary
phase usually applies to integrals in which $T$ appears. We deal always with
billiard examples for which it is convenient to take $\hbar =2m=1,$ and $kL$
is typically large.

The action $S$ generates the classical {\em surface of section map }from 
\begin{equation}
p_s=\partial S/\partial s,\,\,p_{s^{\prime }}=-\partial S/\partial s^{\prime
}.  \label{SSMap}
\end{equation}
Composition of powers of $T$ have intermediate points of stationary phase
determined by this map so in effect, longer and longer orbits can be built
up by iteration of the $T$ operator.

Corresponding to Eq. (\ref{BIM}) is the fundamental equation 
\begin{equation}
\psi (s)=\int ds^{\prime }T(s,s^{\prime }|E)\psi (s^{\prime }).  \label{Tpsi}
\end{equation}
Our new method, for the cases corresponding to special states, solves this
equation directly and quasiclassically, that is, it exploits the stationary
phase approximation to do the integral.

Eq. (\ref{Tpsi}) has solutions only for $E$ values on the quasiclassical
spectrum such that 
\begin{equation}
D(E)=\det (1-T(E))=0  \label{FredD}
\end{equation}
This Fredholm determinant is a well defined version of the {\em dynamical
zeta function}\cite{zeta}, a resummation of the trace formula\cite{Gutz}.
The trace formula itself is given by 
\begin{eqnarray}
d_{osc}(E) &=&\frac{-1}\pi \rm{Im}\frac{d\ln D(E)}{dE}  \nonumber
\label{doscT} \\
\  &=&\frac{-1}\pi \rm{Im}\frac{d\rm{Tr}\ln \left( 1-T(E)\right) }{%
dE}  \nonumber \\
\  &=&\frac{-1}\pi \rm{Im}\frac d{dE}\sum_r\frac 1r\rm{Tr}T^r
\label{doscT}
\end{eqnarray}
Here the density of states, $d(E),$ is expressed as 
\begin{equation}
d(E)=\sum_a\delta (E-E_a)=d_{Weyl}(E)+d_{osc}(E)  \label{D(E)}
\end{equation}
where $d_{Weyl}$ is the smoothed density of states. The traces of powers of $%
T$ are expressed in terms of periodic orbits when the integrals are done in
stationary phase. In the case of the special states, as well as the
integrable case, however, only $r-1$ integrals are well approximated by that
method.

\subsection{Special states and periodic orbits}

We take as an example the slightly sloped billiard described above.
Introduce a reference billiard system consisting of a channel of width $2R,$
i.e. with boundaries at $y=\pm R.$ The actual orbits are compared with the
bouncing ball orbits of this channel.

We want to choose a surface of section such that the nominal periodic orbits
are described by a diagonal element, $T(x,x).$ Such a surface of section is
the upper half of the billiard. It is convenient to label position along it
by the corresponding position $x$ on the upper side of the reference
billiard. [Note that a simple change of coordinates, e.g. $s\rightarrow s(x)$
does not affect the result.]

We now approximate the actual $T(x,x^{\prime })$ in a way valid for $\left|
x-x^{\prime }\right| <<R.$ Thus, we may expand 
\begin{equation}
L(x,x^{\prime })\approx 4R+\frac 1{2R}\left( x-x^{\prime }\right) ^2-\xi
_S(x)-\xi _S(x^{\prime })  \label{LBill}
\end{equation}
where $\xi _S$, given by Eq. (\ref{xiS}) above, is small except in the
endcaps. The first two terms in this formula are an approximation to the
perfect channel, and $\xi _S$ approximates the difference between the sloped
billiard and the channel.

Assuming for the moment that $k\xi _S(x)<<1,$ we can approximate 
\begin{eqnarray}
T(x,x^{\prime }) &\approx &(.)e^{ikR\left( 4+\frac 12\left( \frac{%
x-x^{\prime }}R\right) ^2\right) } \nonumber \\
&\times&e^{-ik\left( \xi _S(x)+\xi _S(x^{\prime
})\right) }  \label{Tapp1} \\
&\approx &(.)e^{ikR\left( 4+\frac 12\left( \frac{x-x^{\prime }}R\right)
^2\right) } \nonumber \\
&\times&\left( 1-ik\left( \xi _S(x)+\xi _S(x^{\prime })\right) \right) .
\label{Tapp}
\end{eqnarray}
This $T$ operator has a form that allows direct solution of Eq. (\ref{Tpsi}%
). Next assume that the solution $\psi (x)$ of Eq. (\ref{Tpsi}) is slowly
varying and expand $\psi (x^{\prime })\approx \psi (x)+\left( x^{\prime
}-x\right) \psi ^{\prime }(x)+\frac 12\left( x^{\prime }-x\right) ^2\psi
^{\prime \prime }(x).$ The rapidly varying exponential in $T$ ensures that
the important contributions to the $x^{\prime }$ integral lie close to $x$,
according to the estimate $\left( x^{\prime }-x\right) ^2\sim R/k.$

One may carry out the integral of Eq. (\ref{Tpsi}) and obtain the equation $%
\psi (x)=\exp (i(4kR-E_mR/2k))\psi (x)$ with the condition that $\psi $
satisfies the Schr\"odinger equation (\ref{Spsi}) with potential $V$ of Eq. (%
\ref{VSlope}), where the replacement $n\pi \rightarrow 2kR$ is made. The
energy eigenvalues $k_{n,m}^2$ are found by insisting that the exponential
is unity. Thus, 
\begin{equation}
4k_{n,m}R-E_mR/2k_{n,m}=2\pi n  \label{energy}
\end{equation}

Since $k\xi _S$ becomes large as $x$ goes into the endcap region, the
approximation of Eq. (\ref{Tapp}) seems to break down. However, the wave
function becomes small in that region. As long as the approximate wave
function continues to be small, the approximation is satisfactory.

It is also possible to extend this result to a parameter regime such that $%
k\xi _S$ becomes of order unity or larger\cite{PNZ}. It is only necessary
that $k\xi _S$ be slowly varying compared with the leading term.

\subsection{Nearly integrable phase space region}

The above formulation can be regarded as defining a phase space region where
the system is nearly integrable. The two arguments of $T$ together with the
surface of section map give an area on the 2-dimensional phase space such
that $T$ is well approximated by Eq. (\ref{Tapp}). This, in turn, is the $T$
operator for an integrable system, characterized by being a function of the
coordinate difference, together with a leading correction. For the case
discussed, the surface of section phase space is approximately $\left|
x\right| \leq a,$ $\left| p\right| \leq \hbar k\left| x-x^{\prime }\right|
/R\approx \hbar \sqrt{k/R}.$ The two-dimensional surface of section phase
space region defines a four dimensional phase space region, from which the
orbits only slowly escape.

A classical orbit at a typical point in this phase space region is not too
different from a nearby classical orbit of the ideal integrable region. Of
course, continuing this orbit for long times allows it to escape from the
region. In the case at hand, almost all such orbits are chaotic with
positive Lyapunov exponents.

However, if the fundamental equation $T\psi =\psi $ has a solution $\psi (x)$
which is large only in the nearly integrable region, there is no need to
consider the very long orbits. That gives another condition on the
perturbation. In terms of the effective potential, $V(x),$ it means that $V$
must become repulsive outside the region. The sign of $V$ depends on the
sign of the perturbation, {\em and} on the sign of the quadratic term in the
integrable part of $T.$ Thus, for bouncing ball states between parallel
sides, a perturbation shortening the bounce path is repulsive, while for
radial bounces between circular sides, a perturbation lengthening the path
is repulsive.

\subsection{Trace formula}

Consider the contribution to the trace formula from the period $1$ orbits,
of Eq. (\ref{D(E)}), $d_{osc,1}=\frac{-1}\pi \rm{Im}\frac d{dE}\rm{%
Tr}T.$ In our case, $\rm{Tr}T=\int dxT(x,x).$ The hard chaos case
usually considered assumes that this integral can be done by stationary
phase, with contributions only from the neighborhood of one or more points.
Using Eq. (\ref{Tapp1}) we see that the stationary phase approximation is
not good unless $\epsilon ka>>1.$ In the case of the opposite strong
inequality, $\epsilon ka<<1$ the $x$ integral gives just a factor of $2a,$
to leading approximation. However, we saw that if $\sqrt{\epsilon }ka>1,$
there are strong effects on the wave functions. Thus, although the wave
functions are sensitive to the slope, the energy level correlations are not
in this parameter regime.

This is a fairly general result. Namely, special states are quite sensitive
to perturbations. The energy level correlations are much less sensitive to
the same perturbation.

\section{More examples}

We briefly give a few more examples. The problems already mentioned can be
solved by the Born-Oppenheimer method. In the examples of this section, the
Born-Oppenheimer method is more limited or more cumbersome, or impossible to
use, but the Bogomolny operator method succeeds.

First, the whispering gallery modes can easily be studied. These modes are
concentrated near the boundary of a sufficiently smooth and sufficiently
convex billiard. The case in which the corresponding classical orbits are
also{\em \ }confined near the boundary is the only one studied in detail in
the literature. Assuming a smooth enough convex billiard with everywhere
positive curvature on the boundary, the appropriate coordinates are $s$, the
distance along the perimeter, and $\rho $, the distance from the perimeter
toward the center of curvature at $s.$ Let $R(s)$ be the radius of curvature
at point $s.$ The Born-Oppenheimer ansatz is $\Psi (r|s)=e^{iks}\Phi (\rho
|s)\psi (s),$ $\Phi (\rho |s)=C(s)J_{\nu (s)}(k(R(s)-\rho ))$ and $\psi
(s)=e^{-ikR(s)f(s)}$. Here $J_\nu $ is a Bessel function of the first kind,
with variable index $\nu $ and $C(s)$ is a slowly varying prefactor,
determined by normalization.

The whispering gallery modes are characterized by large $\nu ,$ slightly
less than $kR.$ It turns out that $\nu =kR(1-f)$ where $f$ is small. One
choice is using the Born-Oppenheimer approach, treating the $s$ variation as
slow compared to $\rho $, [except for the explicit factor $e^{iks}$], and
determining $\nu (s)$ by the condition $J_{\nu (s)}(kR(s))=0.$ [A standard
asymptotic formula expresses $J_\nu $ in terms of an Airy function.]
Alternatively, $\psi $ can be determined as the solution of $T\psi =\psi $
and $\Phi $ can be obtained from that. The result is $f(s)=z_m/2^{\frac
13}(kR(s))^{\frac 23},$ where $z_m$ is a zero of the Airy function.

It turns out that the condition $\left| \partial \Phi (\rho |s)/\partial
\rho \right| >>\left| \partial \Phi (\rho |s)/\partial s\right| $ that the
Born-Oppenheimer approximation is valid is the same as the classical
condition for a caustic. [The quantizing caustic would be at approximately $%
\rho =R(s)f(s).$] The method using the $T$ operator can be generalized,
however, to study cases where there are points of vanishing curvature on the
boundary and caustics do not exist. Associated with a caustic is an
adiabatic invariant. As an orbit passes a zero curvature point, the
adiabatic invariant jumps to a new value. The $T$ operator can account for
this, but the Born-Oppenheimer wave function cannot be used. We show in Fig.
5 a whispering gallery state in a stadium billiard with small short side $%
a=0.05R.$ There are no caustics in this case.

A second example is the slightly sloped trapezoidal billiard of Morse and
Feshbach\cite{MF}, which was recently reconsidered by Kaplan and Heller\cite
{KH}. The remarkable states shown in the next two figures were not
mentioned, however. As shown in Fig. 6, this billiard is almost a square,
with vertical sides, say, $x=0,$ $x=1,$ and horizontal sides $y=1,$ $%
y=\epsilon x,$ where $0<\epsilon <<1.$ The Born-Oppenheimer method may be
used to find states concentrated near $x=0$, similar to the states of the
slightly sloped stadium. However, there are states associated with the $%
(1,1) $ period orbits of the unperturbed square. These orbits are rectangles
making a $45^{\circ }$ angle with the $x-$axis. This may be solved using the
Bogomolny technique by extending the billiard antiperiodically in the $x$
direction, and using as surface of section the upper boundary, $y=1.$ The
effective potential is $V(x)$ $\propto \epsilon k^2\left| x\right| ,$ $%
-1\leq x\leq 1,$ repeated with period $2.$ States of this extended problem
are superimposed to find a solution. A typical case is shown in Fig. 7.

\vspace{2mm}
\begin{figure}[tbp]
{\hspace*{0.7cm}\psfig{figure=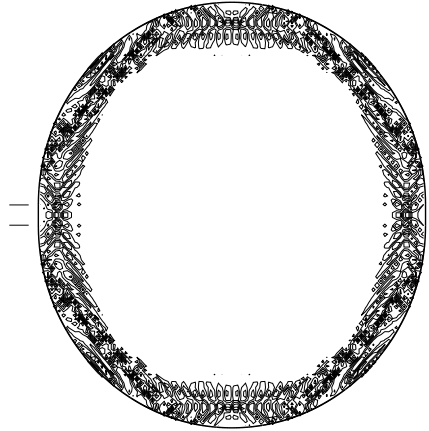,height=6.0cm,width=6.0cm,angle=0}}  
{\vspace*{.13in}}
\caption[ty]{Contour plot of a numerically obtained whispering gallery mode in a
stadium billiard, $a=0.05R.$ There are no caustics in this case, and almost
all classical orbits circulating near the boundary eventually escape. The
distance between the short parallel lines is the length of the straight
side. The wavenumber is $k=242.7611/R.$}
\end{figure}   

A last example is again the almost circular stadium. We now study states
near a higher period orbit. In Fig. 8 we show a numerical state near the
(1,4) almost square orbits in a stadium with side $a/R=0.01.$ We have not
found any convenient coordinates which we can classify as fast and slow, so
the EBO does not work. The $T-$ operator approach is straightforward, but a
little complicated\cite{PNZ}. In the $T-$ operator approximation, there are
two nearly degenerate states. That is, the $T$ operator has inequivalent
solutions of the same energy corresponding to orbits moving clockwise or
counterclockwise. These states do not couple in the $T$ operator
approximation, and the appropriate symmetric combinations have nearly the
same energy.

From this figure it is seen that the states are made up of waves along the
rays of classical angular momentum $l\approx \hbar k/\sqrt{2}$, i.e. along
straight lines whose closest approach to the center is $1/\sqrt{2}.$ A
perfect circle would have a caustic of this radius. Because the stadium is
chaotic, such a caustic does not exist, except as a short time approximation.

\section{Summary}

We have reviewed some of the salient features of special classes of states
which often occur in the systems of interest to quantum chaologists. Perhaps
these cases are common because there is a tendency to construct billiards of
straight lines and circles. The system as a whole may be chaotic, in other
words, almost all orbits of the system may have positive Lyapunov exponents.
However, for some shorter time, a subset of phase space may be close to that
of a set of nonisolated periodic orbits of some integrable reference system,
and this leads to the special states.

\vspace{2mm}
\begin{figure}[tbp]
{\hspace*{0.2cm}\psfig{figure=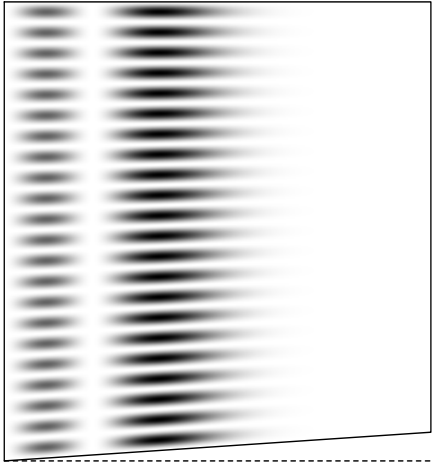,height=6.6cm,width=6cm,angle=0}}  
{\vspace*{.13in}}
\caption[ty]{State along the long side of a nearly square trapezoid. Slope of
bottom is $\epsilon =1/16.$ Quantum numbers are $n=22,$ $m=2.$ In this
figure and the next, the dashed line is the $x-$axis.}
\end{figure}   

These special states have rather striking wavefunctions. As a result, there
are phenomena and even possible applications associated with them. In
particular, the whispering gallery modes have long been known to give rise
to interesting effects. One standard `application' is that special states
often appear in weakly perturbed integrable systems. If these states are to
be avoided, a regular resonant cavity must be constructed much more
precisely than the condition $\delta x<<\lambda ,$ where $\delta x$ is the
deviation from the ideal. Also, because there are a sequence of special
states regularly spaced in energy, the special states often numerically
dominate the trace formula.

Typically, there are two or more scales of variation in connection with the
special states which can be identified. If coordinates can be found such
that one coordinate is fast and the other is slow, standard adiabatic
approximations, such as the Born-Oppenheimer approximation, can be used.

Bogomolny introduced a surface of section transfer operator some time ago as
a means of studying the spectrum, i.e. the trace formula. We have shown how
this operator can also used to find the special eigenstates. The technique
works if the operator can be approximated as a rapidly varying integrable
part, and a more slowly varying correction to it. It is in some ways more
general than the other methods.

\vspace{2mm}
\begin{figure}[tbp]
{\hspace*{1.4cm}\psfig{figure=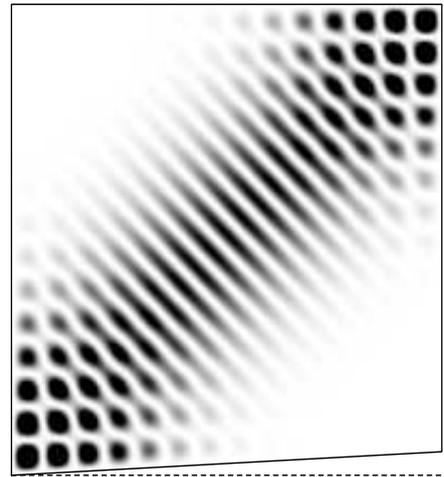,height=6.6cm,width=6.0cm,angle=0}}  
{\vspace*{.13in}}
\caption[ty]{State along the diagonal of the nearly square trapezoid of Fig. 6.
Quantum numbers are $n=27,\,m=1.$}
\end{figure}   

\vspace{2mm}
\begin{figure}[tbp]
{\hspace*{0.2cm}\psfig{figure=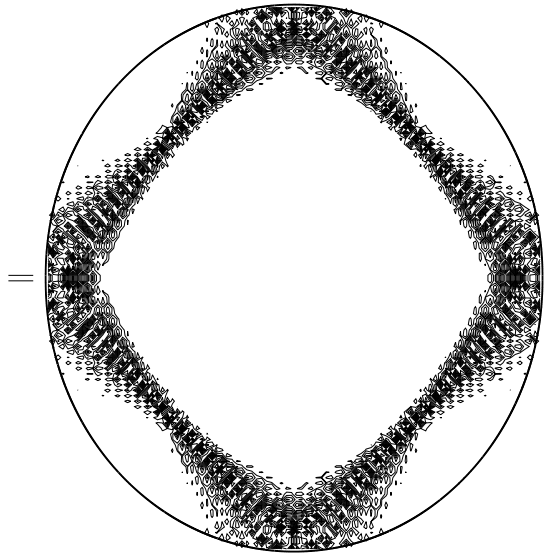,height=7.7cm,width=7.5cm,angle=0}}  
{\vspace*{.13in}}
\caption[ty]{A contour plot of a numerically obtained state near the (1,4)
periodic orbits in a Bunimovich stadium billiard with a short straight side, 
$a/R=0.01.$ The wavelength of the state is $\lambda =2.23969a.$ The parallel
lines show the length of the short side, $2a$.}
\end{figure}   

The special states are typically rather rare, in the sense that they are a
small fraction of all the states in a given (high) energy range. In leading
approximation, they do not couple to the other states. The energies of the
states are predicted to good approximation, relative to the energy spacing
of the given class of states, and even better absolutely. However, the
accuracy is not necessarily good compared with the mean level spacing of all
the levels. Further, it may happen that `accidentally' there is a nonspecial
state with energy very close to that of the special state. Then terms
neglected in our approximation can mix these states. Many phenomena are
independent of such mixing, however.

In this article we gave a number of examples of such special states, several
of which appear for the first time in print. We hope our pictures will tempt
an experimentalist to find some of these states in one of the several
systems, water trays, acoustic plates, microwave and laser cavities, optical
fibers, quantum dots, $\ldots ,$ to which the theory applies.

\section{Acknowledgments}

Supported in part by the United States NSF grant DMR-9625549 and United
States-Israel Binational Science Foundation, grant 99800319. R.N. was
partially supported by the NSF grant DMR98-70681 and the University of
Kentucky.

\end{document}